# Aide et Croissance dans les pays de l'Union Economique et Monétaire Ouest Africaine (UEMOA) : retour sur une relation controversée

Nimonka BAYALE[1]

**Résumé**

*L'objectif de cet article est d'analyser les effets de seuil de l'aide publique au développement sur la croissance économique des pays de la Zone UEMOA. Pour ce faire, l'étude s'est basée sur les données de l'OCDE et du WDI couvrant la période 1980-2015 et a employé un modèle de type Panel Threshold Regression (PTR) à la Hansen pour déterminer par "bootstrap" le seuil d'aide à partir duquel son efficacité est effective. Les résultats de cette méthode appuient fortement l'idée que la relation entre l'aide et la croissance est non linéaire avec un seuil de 12.74 % du PIB moyen de l'Union. Au-dessus de ce seuil, l'effet marginal de l'aide est 0.69 point, toutes choses égale par ailleurs. Ainsi, l'une des principales contributions de cette étude consiste à montrer que les pays de la Zone expriment un besoin de financement qui ne peut être uniquement comblé par l'aide extérieure. Ils devront alors poursuivre leurs efforts à mobiliser et à gérer de façon optimale les ressources internes afin de combler ce besoin et considérer d'aide juste comme une ressource complémentaire.*

**Mots clés** : Aide, *Panel Threshold Regression*, Croissance économique, Zone UEMOA.
**Classification JEL** :  $F_{35}$  $C_{63}$  $C_{26}$  $O_{40}$

# Aid and Growth in West African Economic and Monetary Union (WAEMU) countries : a return back to a controversial relationship

**Abstract**

*The main purpose of this paper is to analyze threshold effects of official development assistance (ODA) on economic growth in WAEMU zone countries. To achieve this, the study is based on OECD and WDI data covering the period 1980-2015 and used Hansen's Panel Threshold Regression (PTR) model to "bootstrap" aid threshold above which its effectiveness is effective. The evidence strongly supports the view that the relationship between aid and economic growth is non-linear with a unique threshold which is 12.74% GDP. Above this value, the marginal effect of aid is 0.69 points, "all things being equal to otherwise". One of the main contribution of this paper is to show that WAEMU countries need investments that could be covered by the foreign aid. This later one should be considered just as a complementary resource. Thus, WEAMU countries should continue to strengthen their efforts in internal resource mobilization in order to fulfil this need.*

**JEL Classification** :  $F_{35}$  $C_{63}$  $C_{26}$  $O_{40}$

**Introduction**

L'aide publique au développement (APD)[2] est l'une des stratégies qui vise à promouvoir la croissance économique dans les pays en développement et contribuer ainsi à y réduire la pauvreté. En effet, depuis plus de cinq (05) décennie, ces pays bénéficent des flux d'aide. Cependant, leur efficacité est loin de faire l'unanimité au sein de la communauté internationale. Certaines études récentes portant sur la relation entre l'aide et la croissance économique dans les pays en développement révèlent que l'aide contribuerait à améliorer la croissance dans ces pays (Stiglitz, 2002 ; Sachs, 2005 ; Dreher et Logman, 2015 ; Ndikumana et Pickbourn, 2016 ; Sothan, 2017) ; alors que d'autres études aboutissent à des résultats absolument controversés. Ces dernières soulignent l'inefficacité de l'aide. Elle n'a donc pas d'effet significatif sur la croissance. Elle contribue parfois même à nuire cette dernière (Friedman, 1958 ; Easterly, 2004 ; Moyo, 2009 ; Mbah et Amassoma, 2014 ; Sraieb, 2016).

Face à cette controverse observée au niveau empirique d'une part et à l'Agenda international 2015-2030 qui insiste sur le doublement de l'aide pour l'atteinte des objectifs cibles de développement (ODD) d'autre part (Nations Unies, 2015), il paraît crucial de tester empiriquement la nature de la relation existant entre l'aide internationale et la croissance économique en s'ancrant dans un espace institutionnel et géographique spécifique, à savoir l'Union Economique et Monétaire Ouest Africaine (UEMOA) qui est un regroupement de huit Etats de l'Ouest africain ayant en commun l'usage de la monnaie CFA. Tel est l'objectif majeur poursuivit dans le présent article. Du point de vue méthodologique, il s'appuie sur l'une des techniques économétriques utilisées dans l'estimation des relations non linéaires : le *« Panel Threshold Regression »* (PTR) à la Hansen (1999).

La suite du travail est structurée de la façon suivante : la première section présente les faits stylisés en rapport avec l'aide internationale au sein de l'UEMOA. Elle effectue également les comparaisons internationales entre les taux de croissance sous régionale et celle de l'Union. La deuxième section fait un état des lieux sur les contributions théoriques et empiriques de l'efficacité de l'aide. La troisième section expose l'analyse économétrique et discute des résultats obtenus. Enfin, la quatrième section conclut et formule les recommandations de politiques économiques.

1. **Les faits stylisés**

La présente section est relative aux faits stylés sur les dynamiques structurelles de l'aide internationale et de la croissance au sein de l'UEMOA pendant ces dernières décennies.

**1.1. L'aide publique au développement dans les Etats de l'UEMOA**

L'évolution de l'aide publique au développement (évaluée en % du PIB) au sein des économies de l'UEMOA est illustrée par le graphique ci-dessous (graph 1). En effet, celui-ci indique que pour l'ensemble des Etats de l'Union, la variation moyenne semble se stabiliser autour de 12 %. Si l'on entre dans les détails, on constate que depuis les années 80, c'est en Guinée Bissau qui a atteint des niveaux d'aide les plus élevés, avec un pic de 68.71 % en 1994. Sur la période 1980-2015, il apparait que deux (02) pays ont reçu des quantités d'aide relativement faibles. Ces pays sont la Côte d'Ivoire et le Togo, respectivement 5 % et 8 % en moyenne. Cette situation pourrait s'expliquer par le contexte économique et institutionnel qui ont régné dans ces pays. La Côte d'Ivoire a l'atout d'être l'une des économies les plus diversifiées de l'Union. Elle apparait donc comme un pays moins dépendant de l'aide extérieure. Pour ce qui concerne le Togo, il a été sevré de l'aide pendant plus d'une dizaine d'années (1993 – 2007).

---

[2] - Dans cet article, nous traitons de l'aide globale. De façon classique, son contenu regroupe : l'aide aux projets et programmes, la coopération technique, l'aide humanitaire et la remise de dette. Généralement, les deux premières composantes constituent la part la plus importante. En 2014 par exemple elles ont représenté près de 60 % de l'APD totale nette (OCDE, 2016).

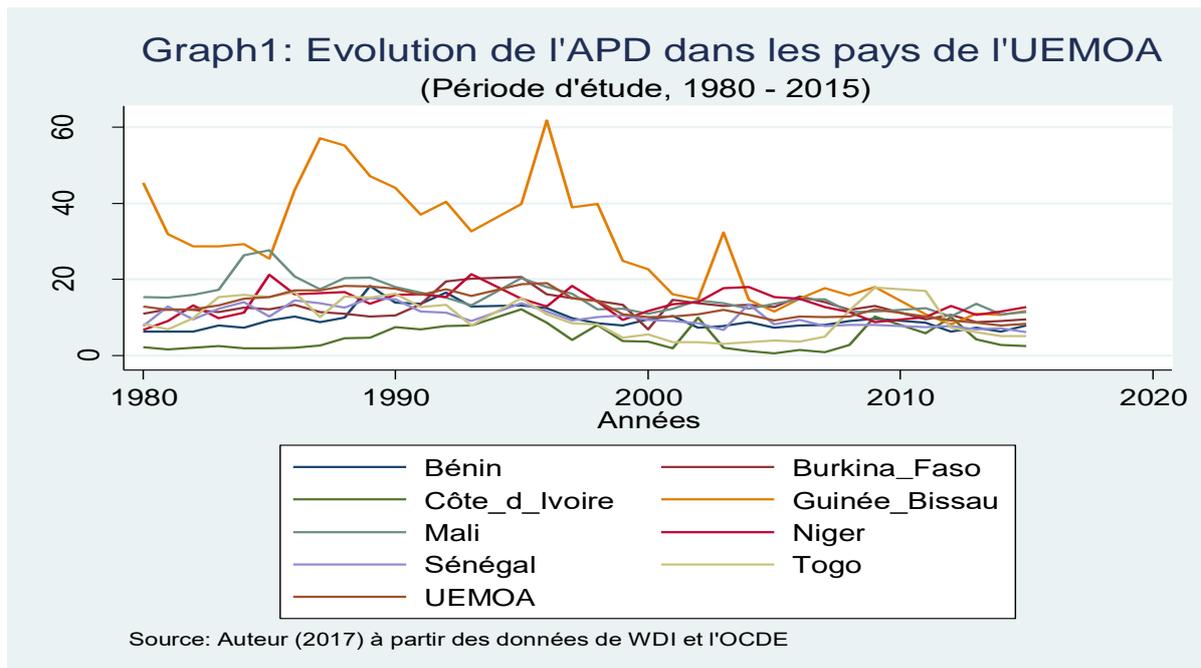

Globalement, lorsqu'on isole la Guinée Bissau, on observe une tendance relativement stable pour l'ensemble de l'échantillon. Cependant, l'aide au Bénin, au Mali et au Niger a connu une évolution en dent de scie avant de se stabiliser autour des années 2000, à l'exception des 14 % à partir de 2013 pour le Mali et des 18 % en 2004 pour le Niger. Dans l'ensemble, on a observé une certaine convergence du niveau de l'aide estimée en % du PIB pour la plupart des pays de l'UEMOA à partir de 2004 où l'aide oscille entre 3 % et 18 %.

### 1.2. Evolution récente de la croissance du PIB dans l'UEMOA et les comparaisons internationales

Au cours de ces dernière années, la zone UEMOA est connue comme l'une des régions où la croissance a été solide et forte avec la Côte d'Ivoire et le Sénégal qui restent les économies les plus pondérées de l'Union (FMI, 2017). Cependant, elle a connu des périodes pendant lesquelles la croissance n'a pas été reluisante. La période relative aux années 80 et qui coïncidait avec les Programmes d'Ajustement Structurel (PAS) en fait partie. En effet, le graphique (graph2) présenté ci-dessous indique que sur la période 1980-2015, le taux de croissance moyen du PIB de la zone s'établit à 3.28 %. Comparativement aux autres régions et sous régions, ce taux est relativement plus élevé que ceux de l'Afrique subsaharienne (3.14 %), de l'Amérique latine et caraïbe (2.58 %) et de l'Union Européenne (1.90 %). Par contre, il reste nettement inférieur à celui des pays d'Asie du Sud-Est (7.98 %). Par ailleurs, dans l'UEMOA, les moyennes les plus fortes sont observées au Burkina Faso (4.14%), Bénin (4.05%) et Mali (3.82%). Par contre, la Guinée Bissau (2.39%) où l'incertitude politique demeure un frein réel au décollage économique du pays, la Côte d'Ivoire (2.57%) et le Togo (2.79) pour leurs multiples crises socio-politiques, se sont révélés comme des pays dans lesquels la croissance moyenne a été relativement faible.

Graph 2 : Taux de croissance du PIB dans l'UEMOA et comparaisons internationales, valeurs moyennes de la période 1980 – 2015

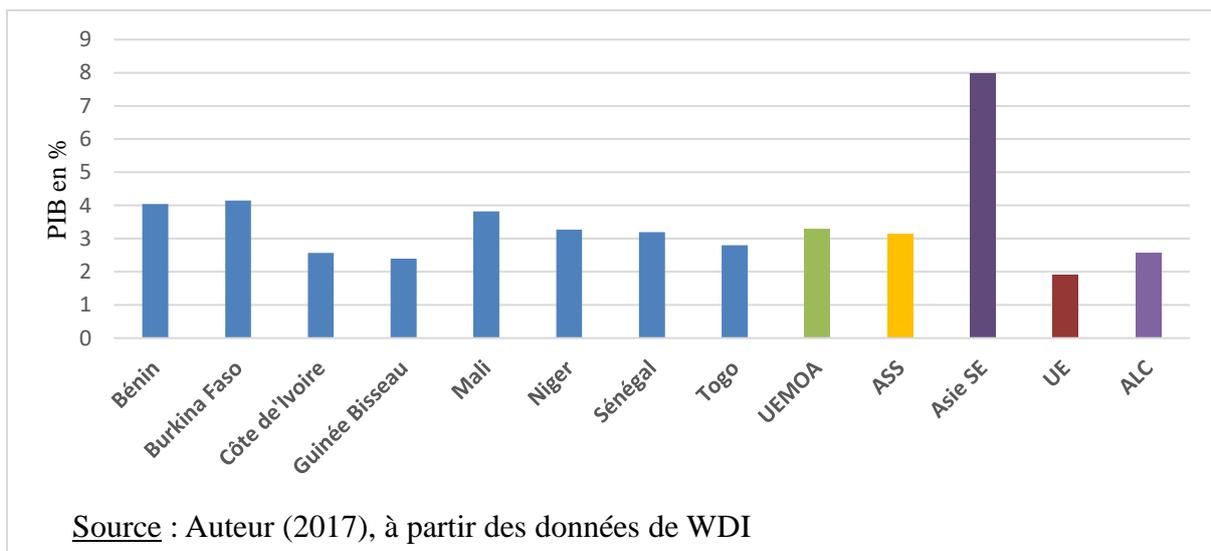

Source : Auteur (2017), à partir des données de WDI

De plus, sur la même période, le Sénégal (3.21%) et le Niger (3.28%) ont enregistré une croissance relativement stable surtout lorsqu'on relativise le choc alimentaire des années 1984 – 1985 au Niger qui a conduit le pays à enregistrer un taux croissance de -16.82%. D'autres chocs de nature socio-politique ont conduit le Togo (en 1993) et la Guinée Bissau (en 1998) à enregistrer -15.09% et -28.09% respectivement (valeurs les plus élevées en valeur absolue). Néanmoins, les perspectives économiques récentes des pays l'Union laissent envisager une certaine convergence économique (FMI, 2017). Comme planifié, la section suivante fixe l'ancrage théorique ainsi que les contributions empiriques sur la relation Aide et croissance.

## 2. Revue de littérature : de la théorie aux travaux empiriques

L'importance théorique de l'aide publique au développement pour une économie en besoin de financement remonte aux travaux sur le *« big push »* de Rosenstein-Rodan (1943, 1961) pour qui un apport massif de capitaux extérieurs pouvait permettre aux pays en développement de financer leurs investissements et faire face aux problèmes primaires liés au décollage économique. Toutefois, dans la littérature économique, certains auteurs ont montré que l'aide publique existait bien avant, avec notamment l'aide acheminée par les gouvernements britannique et français vers leurs colonies dans les années 1930 et 1940 (Riddell, 2007). Pour d'autres, le Plan Marshall lancé en 1947 pour la reconstruction de l'Europe de l'après-guerre, l'adoption de la Charte des Nations unies en 1945 et surtout la Déclaration des droits de l'Homme en 1948, ont marqué une étape importante dans ce qui pourrait être considéré comme les principes fondateurs de l'aide au sens moderne du terme (Jacquet et Naudet, 2006).

De toutes ces sources, c'est le cadre théorique du modèle Harrod-Domar (Easterly 1999, 2001) qui a constitué l'ossature intellectuelle sur les éléments instigateurs de la politique d'aide internationale. Vue comme une solution optimale pour des pays se trouvant dans une situation que Solow (1965) a appelé plus tard le *« piège du sous-développement »,* l'aide internationale a également joué un rôle important dans les modèles à *« double déficits »* de Hollis Chenery et Alan Strout (1966). Cette variante du modèle détermine le manque de financement de l'économie à partir de deux déficits : le déficit d'épargne et le déficit au niveau des échanges en capital avec l'extérieur.

Du point de vue empirique, le champ d'analyse de l'efficacité de l'aide au développement, déjà exploré dans les années 1970, a connu véritablement un regain important au début des années 1990. La plupart des études sur l'aide ont cherché à identifier son impact sur la croissance économique, sous-entendu que cette dernière s'accompagnera de progrès sociaux. Cependant, les résultats, dans leur ensemble, prêtent à controverse. Nous pouvons les classer en deux (02) tendances : d'une tendance optimiste à une tendance pessimiste sur l'efficacité de l'aide.

- **De l'optimisme sur l'efficacité de l'aide…**

Jusqu'en 2008, la littérature sur l'efficacité de l'aide contenait au moins environ cent (100) articles dont soixante et huit (68) fournissaient des résultats favorables à l'effet de l'aide sur la croissance économique (Doucouliagos et Paldam, 2008). En effet, commençons par les études de Burnside et Dollar (1997, 2000) qui constituent les analyses de référence en rapport avec l'efficacité de l'aide dans les pays en développement. Pendant que l'aide traversait une crise de légitimité, ces auteurs démontrent que l'efficacité de l'aide est conditionnée par les bonnes politiques économiques et la qualité des institutions dans les pays bénéficiaires. Les conclusions de Burnside et Dollar (1997) seront reprises et défendues par la Banque Mondiale (1998), Dalgaard et Hansen (2001), Lensink et White (2001) avec des implications de politiques économiques parmi lesquelles, ils postulent que si l'aide est plus efficace dans un *bon* environnement macroéconomique. En conséquence, elle devrait cibler les pays pauvres ayant adopté de « *bonnes politiques économiques[3]* ». Cette conclusion fut récemment partagée par Fiodendji et Evlo (2013) qui, dans un contexte de la CEDEAO, ont conditionné l'efficacité de l'aide par les politiques économique et la qualité des institutions. Pour Guillaumont et Chauvet (2001, 2002), les termes de l'échange et même les aspects climatiques pourraient apparaitre parmi les facteurs qui conditionneraient l'efficacité de l'aide.

Précédemment aux travaux de Burnside et Dollar (1997 2000), les résultats empiriques de Dowling et Heimenz (1982) et Rana (1987) confirmaient déjà l'effet significativement positif

---

[3] - Ils ont utilisé un indicateur composite de politiques économiques qui prend en compte l'équilibre budgétaire, l'ouverture commercial et la maîtrise de l'inflation pour juger que la qualité des politiques économiques. Cet indicateur est obtenu à partir de l'équation suivante : Pol eco = 1,28 + 6,85 ∗ surplus bugétaire − 1,40 ∗ inflation + 2,16 ∗ Politique budgétaire.

de l'aide sur la croissance au sein de la région asiatique. Suivant cette logique, Stiglitz (2002), Stern (2002) et Sachs (2005) ont également défendu l'idée que l'aide a, en général, un effet positif sur les pays à faible revenu, en dépit de quelques échecs remarquables. Ils en veulent pour exemple la Corée, le Botswana et l'Indonésie ou encore, de la Tanzanie et du Mozambique. Une autre preuve de l'efficacité de l'aide s'est traduite, disent-ils, par le recul des indicateurs de pauvreté et la progression rapide des indicateurs d'éducation et de santé ces dernières décennies suite à la généralisation de l'aide surtout en Afrique. D'ailleurs, les études contemporaines sur l'efficacité de l'aide dans le contexte de l'Afrique Subsaharienne, les analyses de Arndt et *al*. (2010), de Galiani et *al*., (2014), de Scott (2015), de Dreher et Lohmann (2015), de Aboubacar et *al*., (2015), de Civelli et *al*. (2017) et de Sothan (2017) arrivent à la même conclusion concernant l'effet positif de l'aide sur la croissance économique.

- **… au pessimisme sur l'efficacité de l'aide**

A l'opposé de ces travaux qui mettent en vogue l'efficacité de l'aide, Friedman (1958) et Bauer (1972), dès les premières décennies de l'après-guerre, se sont montrés très critiques vis-à-vis de l'aide publique au développement en évoquant son inefficacité. Ils soulignent que l'aide a un effet négatif sur la croissance des pays en développement en ce sens qu'elle encourageait la mauvaise gestion et ne bénéficiait qu'à une certaine oligarchie des pays en développement au détriment des populations (Mosley et *al*., 1987). L'Haïti et la Somalie furent cités comme exemples. Ainsi, le détournement de l'aide de leur objectif à des fins personnelles et extra économiques constitue l'une des principales causes de son inefficacité (Boone, 1996 ; Severino et Charnoz, 2003). Dans ce même ordre d'idées, Elliot Berg (1997) avait déjà détecté des effets pervers de l'aide sur les décisions des gouvernants des pays bénéficiaires. C'est le cas typique du Nigéria (Mbah et Amassoma, 2014).

De toutes ces analyses, le livre qui a particulièrement fait *l'effet de bombe* concernant l'inefficacité de l'aide est celui de Moyo (2009) intitulé « *Dead aid : why aid is not working and how there is a better way for Africa* ». L'auteur décrit les effets pervers de l'aide qu'il considère comme l'une des principales sources des distorsions socioéconomiques observées sur le continent africain. Elle programme le sevrage progressif des pays africains de cette aide sur une période de 5 à 10 ans. Les analyses de Rajan et Subramanian (2008, 2010), de Temple et Van de Sijpe (2014) et de Dreher et Langlotz (2015) sont venues en renfort à ces conclusions puisqu'elles aboutissent également à une absence d'effet positif de l'aide sur la croissance économique des pays en développement. Toutefois, l'idée du *syndrome hollandais* est évoquée pour expliquer cette non significativité. Parfois, on s'interroge sur non seulement la qualité des politiques économiques menées comme facteur d'efficacité de l'aide, mais également ce qu'elle finance (Hansen et Tarp, 2000 et 2001 et Guillaumont et Chauvet, 2001). L'aide alimentaire par exemple est contre-productive alors que le financement des infrastructures économiques impact positivement la croissance économique à moyen terme selon leurs analyses.

Par extension, certains auteurs analysant l'efficacité de l'aide ont affirmé que l'impact dépendait des pratiques ou des motivations du donateur (Djankov et *al*., 2009, Kilby et Dreher, 2010 ; Dreher et *al*., 2014), de la forme et du volume de l'aide, avec l'introduction de la notion d'aide seuil (Boone, 1996 ; Clemens et *al*., 2004 ; Kalyvitis et *al*. (2012), Fiodendji et Evlo, 2013 ; Alia et Anago, 2014). D'autres expliquent clairement que l'un des facteurs qui freine l'efficacité de l'aide est sa fongibilité (Devarajan et al., 1999 ; Masud et Yontcheva, 2005). C'est pourquoi, ils privilégient la pratique de l'aide projet. Cette recommandation requiert l'assentiment de Azam et Laffont (2003) qui, s'inspirant du *« paradoxe micro-macro »* de l'aide décrit par Mosley (1986), démontrent que l'efficacité de l'aide peut être atteinte grâce à l'aide projet marquée par l'existence d'une coordination entre les gouvernements et les ONG locales.

En définitive, il existe donc une incertitude sur les effets de l'aide sur la croissance puisque la capacité de l'aide extérieure à accélérer la croissance dans le pays d'accueil n'est pas fermement soutenue. La plupart de ces études qui explorent presque toutes l'impact de l'aide sur les taux de croissance partagent les mêmes faiblesses méthodologiques, le premier problème étant le mauvais traitement de l'endogénéité de l'aide et à l'homogénéité des échantillons d'étude (Temple, 2010 ; Kahneman et Deaton, 2010) et le second étant lié à l'ignorance structurelle de l'existence d'une relation non linéaire entre l'aide et la croissance (Fiodendji et Evlo, 2013).

### 3. Analyse économétrique
### 3.1. Le modèle à effet de seuil à la Hansen

Pour le besoin de notre étude, un modèle à effet de seuil de type « *Panel Threshold Regression* » (PTR) initialement introduite par Hansen (1999) et repris plus tard par Im, Pesaran et Shin (2003) pour caractériser la non-linéarité d'une relation liant deux (02) ou plusieurs variables dans un modèle de régression a été adopté. La spécification générale du modèle de seuil prend la forme suivante :

$$y_{it} = \mu_i + \sum_{k=0}^{K-1} \beta_{K+1} x_{it} \, \mathbb{I}(\gamma_k < q_{it} \leq \gamma_{k+1}) + \beta_{K+1} x_{it} \, \mathbb{I}(\gamma_K < q_{it} \leq \gamma_{K+1}) + \varepsilon_{it} \ (1)$$

Où l'indice $i$ réfère à la dimension individuelle avec $(1 \leq i \leq N)$ et $t$ la dimension temporelle $(1 \leq t \leq T)$. $\mu_i$ est l'effet fixe spécifique pour chaque pays et $\varepsilon_{it}$ le terme d'erreur est de ce point de vue indépendantes et identiquement distribuées ($iid$) de moyenne nulle et de variance $\sigma_\varepsilon^2$ finie $[iid\,(0, \sigma_\varepsilon^2)]$.

Dans ce modèle, le mécanisme de transition est modélisé à l'aide d'une fonction indicatrice $\mathbb{I}(.)$ qui prend la valeur 1 si la contrainte entre parenthèse est respectée, et zéro sinon. Elle est définie par la variable de seuil $q_{it}$ et de paramètre de seuil $\gamma$. $y_{it}$ est la variable dépendante et $x_{it}$ est le vecteur des variables explicatives. On note aussi que $\gamma_0 = -\infty$ et $\gamma_{K+1} = +\infty$. L'équation (1) nous permet d'obtenir de $K$ valeurs de seuil et de $(K+1)$ régimes. Au niveau de chaque régime, l'effet marginal de $x_{it}$ ($\beta_k$) sur $y_{it}$ peut alors varier.

Toutefois, suivant l'analyse de Gong et Zou (2001) et de Bick et Nautz (2008), nous considérons un discriminateur constant qui n'est pas spécifiquement individuel mais capture statistiquement l'effet commun pour toute la dimension temporelle. Le fait d'ignorer ($\delta_k$) dans le modèle peut toutefois conduire à des estimations biaisées des valeurs seuil et par conséquent, l'impact des effets marginaux correspondants. Sur cette base, l'équation (1) devient :

$$y_{it} = \mu_i + \sum_{k=0}^{K-1}(\beta_{K+1} + \delta_{k+1})\, x_{it}\, \mathbb{I}\left(\gamma_k < q_{it} \leq \gamma_{k+1}\right) + \beta_{K+1} x_{it}\, \mathbb{I}\left(\gamma_K < q_{it} \leq \gamma_{K+1}\right) + \varepsilon_{it} \quad (2)$$

Où la différence entre les interceptions des régimes est représentée par $\delta_k$. L'équation (2) suppose que $\delta_k$ reste le même pour toutes les sections (la dimension temporelle) mais n'est pas spécifiquement individuel.

Concernant la méthode d'estimation, il pourrait se poser deux problèmes : l'estimation des effets individuels qui sont communs aux différents régimes d'une part et de celle des coefficients de pentes et des paramètres de seuils. Ces derniers rendent impossible l'estimation directe par *Ordinary Least Squares (OLS),* les variables explicatives dépendant de ces paramètres de seuils. Dans cette logique, la démarche d'estimation doit alors s'effectuer, comme le conseille Bai (1997) et Hurlin et *al*., (2008), en transformant le modèle (2) de la manière suivante :

$$\tilde{y}_{it} = \tilde{\mu}_i + \sum_{k=0}^{K-1}(\beta_{K+1} + \delta_{k+1})\, \tilde{x}_{it}\, \mathbb{I}\left(\gamma_k < q_{it} \leq \gamma_{k+1}\right) + \beta_{K+1} \tilde{x}_{it}\, \mathbb{I}\left(\gamma_K < q_{it} \leq \gamma_{K+1}\right) + \varepsilon_{it} \quad (3)$$

Où $\tilde{y}_{it} = y_{it} + \bar{y}_{it}$ avec $\bar{y}_{it} = \frac{1}{T}\sum_{t=1}^{T} y_{it}$ ; $\tilde{u}_{it} = u_{it} + \bar{u}_{it}$ avec $\bar{u}_{it} = \frac{1}{T}\sum_{t=1}^{T} u_{it}$ et $\tilde{x}_{it}(\gamma) = x_{it}(\gamma) + \bar{x}_{it}(\gamma)$ avec $\bar{x}_{it}(\gamma) = \frac{1}{T}\sum_{t=1}^{T} x_{it}(\gamma)$

Une fois que les effets individuels fixes éliminés, la démarche consiste à appliquer les *Moindres Carrés Séquentiels (MCS)*. En effet, pour des seuils fixés, il est possible d'estimer les coefficients de pentes $\beta$. Ainsi, on estime d'abord $\hat{\beta}(\gamma)$ comme suit :

$$\hat{\beta}(\gamma) = \left[\sum_{i=1}^{N}\sum_{t=1}^{T} \tilde{x}'_{it}(\gamma) \times \tilde{x}_{it}(\gamma)\right]^{-1} \left[\sum_{i=1}^{N}\sum_{t=1}^{T} \tilde{x}'_{it}(\gamma) \times \tilde{y}_{it}(\gamma)\right] \quad (4)$$

Ensuite, on en déduit la somme des carrés résidus (SCR) :

$$SCR(\gamma) = \sum_{i=1}^{N}\sum_{t=1}^{T}(u_{it}^{2*}) = \sum_{t=1}^{N}\sum_{t=1}^{T}(\tilde{y}_{it} - \hat{\beta}'(\gamma) x_{it}(\gamma))^2 \quad (5)$$

Cette démarche doit être répétée pour l'ensemble des valeurs de seuils possibles comprises dans un intervalle $\Omega$, qui est défini de façon à garantir un nombre minimum d'observations dans chaque régime. Chan (1993) et Hansen (1999) recommandent de retenir comme estimateurs optimaux des paramètres de seuils $\hat{\gamma} = (\hat{\gamma}_1, \ldots\ldots\ldots, \hat{\gamma}_{k+1})$, ce qui minimisent la somme des carrés des résidus :

$$\hat{\gamma} = \arg\min_{\gamma \in \Omega} SCR(\gamma) \quad (6)$$

Les coefficients de pentes $\beta'(\gamma)$ sont alors à nouveau obtenus à l'aide du *Ordinary Least Squares (OLS)* calculées en $\hat{\gamma}$. Ce qui rend alors possible la déduction de la variance empirique des résidus :

$$\hat{\sigma}^2 = \sum_{i=1}^{N}\sum_{t=1}^{T}\frac{1}{n(T-1)}\hat{u}_{it}^* \hat{u}_{it}^* = \frac{1}{n(T-1)}S(\hat{\gamma}) \qquad (7)$$

Après ce traitement, nous introduisons à présent un régime d'intersection dans un modèle de seuil pour éliminer l'effet spécifique individuel avec des effets fixes standards en transformant les coefficients de pente $\beta_k$ *et* $\beta_{k+1}$ en $\beta_1$ et $\beta_2$ (cas de deux régimes). Ainsi, cette forme particulière d'un modèle à effet de seuil à Hansen à deux (02) régimes s'écrit :

$$y_{it} = \mu_i + \beta_1 x_{it}\mathbb{I}(q_{it} \leq \gamma) + \delta_1 \mathbb{I}(q_{it} \leq \gamma) + \beta_2 x_{it}\mathbb{I}(q_{it} > \gamma) + \epsilon_{it} \qquad (8)$$

Dans cette équation, $\mathbb{I}(q_{it} \leq \gamma)$ représente le régime d'intersection. Cette formulation de l'équation (8) suppose que la différence entre les intersections de régimes, représenté par $\delta_1$, n'est pas spécifiquement individuel mais plutôt la même pour toutes les sections. Alors, les estimations de pente pour chaque régime sont identiques à celles d'une régression en utilisant uniquement des observations du régime qui reflète l'orthogonalité des variables explicatives $\mathbb{I}(x_i \leq x_m)$ et $x_i\mathbb{I}(x_i > x_m)$. Elles peuvent être obtenues par la méthode *OLS*. Mais si elles sont biaisées, elles entrainent d'autres conséquences dans le modèle de seuil des données de panels. C'est pourquoi pour obtenir des estimateurs de $\beta_1$ *et* $\beta_2$ *best linear unbiaised estimateur (BLUE),* il est nécessaire d'étendre la configuration du modèle initial en introduisant la variable retardée de $y_{it}$ en référence à Hansen et Caner (2004), à Drukker et *al.*, (2005) et à Vinayagathasan (2013). Le modèle estimable devient :

$$y_{it} = \mu_i + \theta_1 y_{i,t-1} + \beta_1 x_{it}\mathbb{I}(q_{it} \leq \gamma) + \delta_1 \mathbb{I}(q_{it} \leq \gamma) + \beta_2 x_{it}\mathbb{I}(q_{it} > \gamma) + \epsilon_{it} \qquad (9)$$

Dans cette équation (9), $y_{i,t-1}$ est la variable retardée d'une période de la variable dépendante $y_{it}$. Pour un contrôle rigoureux de tout biais, il est préférable que cette équation soit estimée par GMM ou par la méthode instrumentale du fait l'endogénéité avérée de l'aide publique au développement, lorsque tous les tests pré-estimations sont concluants (Hansen et Caner, 2004).

Dans cette logique, l'hypothèse nulle pour tester la signification du seuil doit être étendue par $\delta_1 = 0$ d'une part, et la dérivation de la distribution asymptotique de l'estimation de seuil repose maintenant sur l'hypothèse technique additionnelle suivante : $\delta_1 \to 0$ lorsque $N \to \infty$ d'autre part. Cela signifie que la différence dans les interceptions entre les deux régimes est « *minimisable* » par rapport à la taille de l'échantillon qui est complètement analogue à l'hypothèse concernant les coefficients de pente. C'est pourquoi à l'annexe de son article Hansen (1999) démontre théoriquement que les expressions : $\theta' = [(\beta_2 - \beta_1)' - \delta_1]$ et $z_{it} = (x_{it}' 1)C$ pourraient être prises en compte comme régresseurs supplémentaires du régime d'intersection (Im, Pesaran et Shin, 2003).

### 3.2. Inférence dans le modèle *PTR*
- **Le test de linéarité**

C'est un test crucial qui consiste à prouver si l'effet de seuils est statistiquement significatif et réciproquement de montrer que la relation liant les variables explicatives à la variable expliquée peut être représentée à l'aide d'un modèle à changements de régimes. Pour ce faire, on construit un test d'hypothèse nulle de linéarité contre l'alternative d'un modèle à transition brutale avec un unique seuil. Plus précisément, ce test consiste à tester l'égalité des coefficients des différents régimes. Dans l'équation (9), l'absence d'effet de seuil est représentée par l'hypothèse suivante :

$$H_0: \beta_1 = \beta_2 \ contre \ H_0: \beta_1 \neq \beta_2$$

Le test est alors construit en considérant le seuil comme étant fixé à sa valeur estimée. Il est ainsi possible d'utiliser les statistiques de tests usuels telles que celui de Fisher :

$$F_1 = \frac{S_0 - S_1(\hat{\gamma}_1)}{\hat{\sigma}^2} \quad \text{où} \quad \hat{\sigma}^2 = \frac{1}{N(T-1)} S_1(\hat{\gamma}_1) \qquad (10)$$

Où $S_0$ est la somme des carrés des résidus du modèle linéaire et $S_1(\hat{\gamma}_1)$ la somme des carrés des résidus du modèle à un seuil. Toutefois, comme l'estimateur du paramètre de seuil est obtenu par maximisation de la fonction de vraisemblance des observations, la distribution des statistiques de tests n'est pas connue. La résolution passe par la méthodologie de Hansen (1996).

- **Le test de détermination du nombre de régimes**

Ce test s'applique en cas de présence d'un effet de seuils avérée. Sa procédure est similaire à celle utilisée pour tester la linéarité. Par exemple, pour tester si le modèle possède deux régimes (hypothèse nulle $H_0: \beta_2 = 0$), ou au minimum trois régimes (hypothèse alternative $H_1: \beta_2 \neq 0$), le test de Fisher suivant est appliqué :

$$F_2 = \frac{S_1(\hat{\gamma}_1^*) - S_2(\hat{\gamma}_1^*, \hat{\gamma}_2^*)}{\hat{\sigma}^2} \qquad (11)$$

où $S_2$ est la somme des carrés des résidus du modèle à trois régimes. L'hypothèse nulle d'un seuil unique est rejetée en faveur d'au minimum deux, si la valeur de $F_2$ est supérieure aux valeurs critiques simulées par *bootstrap*.

En cas de rejet de l'hypothèse nulle d'un modèle à un seuil, la démarche de détermination du nombre de régimes se poursuit. Il est alors nécessaire de tester l'hypothèse nulle de deux seuils ($H_0: \beta_3 = 0$) contre un modèle contenant au minimum trois ($H_1: \beta_3 \neq 0$). Le test de Fisher $F_3$ correspondant se présente dans l'équation (11) comme suit :

$$F_3 = \frac{S_2(\hat{\gamma}_1^*, \hat{\gamma}_2^*) - S_3(\gamma_1^*, \gamma_2^*, \hat{\gamma}_3^*)}{\hat{\sigma}^2} \qquad (12)$$

où $S_3$ est la somme des carrés des résidus du modèle à quatre (04) régimes. En cas de rejet de l'hypothèse nulle, la spécification doit contenir au minimum quatre régimes, et la démarche devrait être poursuivie jusqu'au non rejet de l'hypothèse nulle[4].

---

[4] - Dans la pratique, on s'arrête le plus souvent à quatre. La solution à privilégier pour pouvoir en considérer un plus grand nombre est d'utiliser un mécanisme de transition lisse et non plus brutale (Hurlin et Fouquau, 2008).

- **Intervalle de confiance sur le seuil**

Lorsque l'effet de seuils est avéré et que le nombre de régimes est déterminé, Chan (1993) et Hansen (1999) montrent que les seuils obtenus $\hat{\gamma}$ sont des estimateurs convergents des vraies valeurs et que la distribution asymptotique de ceux-ci est non standard. Pour estimer leur intervalle de confiance, il est alors nécessaire de former une région de non rejet en se servant des tests de Fisher qui viennent d'être présentés lors des tests de non linéarité. Ainsi, dans une représentation PTR à $K+1$ régimes, le test de Fisher revient à tester $H_0 : \gamma_j = \gamma_0$ contre $H_0 : \gamma_j = \gamma_0$ à partir de la statistique de test suivante :

$$LR(\gamma) = \frac{S(\gamma_j/\gamma_1, \ldots, \gamma_{j-1}; \gamma_{j+1}, \ldots, \gamma_m) - S(\hat{\gamma}_j/\gamma_1, \ldots, \gamma_{j-1}; \gamma_{j+1}, \ldots, \gamma_m)}{\hat{\sigma}^2} \qquad (13)$$

où $\gamma_0$ est la vraie valeur du seuil, $\gamma_j$ représente le seuil sur lequel l'intervalle de confiance est créé et $S(\gamma_j)$ est la somme des carrés des résidus obtenus en $\gamma_j$ conditionnellement aux autres seuils. Sachant que l'hypothèse nulle est rejetée pour des fortes valeurs de $LR_1(\gamma_0)$, l'intervalle de confiance à $(1 - \alpha)\%$ est donc la zone de « *non rejet* », ou autrement dit l'ensemble des valeurs de $\gamma_j$ pour lesquels $LR_1(\gamma) \leq \gamma(\alpha)$ où $\gamma(\alpha)$ représente les valeurs critiques du test associées à un risque de première espèce de α%. Pour obtenir ces derniers, Hansen (1999) considère les valeurs critiques suivantes :

$$\gamma(\alpha) = -2 \log \left(1 - \sqrt{1-\alpha}\right) \qquad (14)$$

La conclusion théorique du test montre que l'hypothèse nulle $H_0 : \gamma_j = \gamma_0$ est rejetée pour un risque $\alpha$ si la valeur de $LR_1(\gamma)$ dépasse $\gamma(\alpha)$.

### 3.3. Spécification du modèle empirique

Comme définit plus haut, nous appliquons le modèle a seuil modifié sur données de panel dans l'espace UEMOA afin d'analyser l'effet de l'aide sur la croissance économique. Pour ce faire, nous considérons la spécification empirique suivante :

$$\begin{aligned} TCPIB_{i,t} = \mu_{i,t} + \theta_1 TCPIB_{i,t-1} + \beta_1 APD_{i,t}\mathbb{I}(APD_{i,t} \leq \gamma) + \delta_1 \mathbb{I}(APD_{i,t} \leq \gamma) \\ + \beta_2 APD_{i,t}\mathbb{I}(APD_{i,t} > \gamma) + \alpha_i X_{i,t} + \varepsilon_{i,t} \end{aligned} \qquad (15)$$

$i = indice\ pays\ et\ t = indice\ temporel$

Dans cette équation (15), de la gauche vers la droite, sont libellés :
- $TCPIB_{i,t}$ : le taux de croissance du PIB. C'est la variable expliquée par un ensemble de variables dont les valeurs retardées d'une période de la variable dépendante ;
- $APD_{i,t}$ : l'aide publique au développement (globale) reçue par chacun des pays de l'échantillon de l'étude. Elle est rapportée au PIB (en % du PIB).

Les fonctions indicatrices $\mathbb{I}(APD_{i,t} \leq \gamma)$ et $\mathbb{I}(APD_{i,t} > \gamma)$ sont susceptibles de prendre des valeurs 1 si le terme entre la parenthèse est vrai et 0 si la condition en parenthèse n'est pas vérifiée. On présume que $\beta_1 \leq 0$ et $\beta_2 > 0$ car en théorie une aide faible enfonce les pays dans le *piège du sous-développement* alors qu'une aide forte à la manière du *big push* affectera positivement le $TCPIB$ (Rosenstein-Rodan, 1961).

- $X_{i,t}$ est une matrice de variables de contrôle susceptibles d'expliquer le taux de croissance de l'économie. Salai-i-Martin (1997) a identifié plus soixante (60) variables ayant un effet significatif sur la croissance économique dans au moins une équation de régression. Toutefois, dans la présente analyse, nous limitons leur nombre[5] pour apprécier au mieux l'influence de notre variable d'intérêt.

Cette matrice inclut notamment : *un indice de qualité institutionnelle* (INST) construit à partir d'une moyenne pondérée de quatre (04) variables institutionnelles d'*International Country Risk Guide (ICRG)* : le contrôle de la corruption, la stabilité du gouvernement (politique), la responsabilité démocratique et la qualité bureaucratique. En effet, un cadre institutionnel sain et solide est favorable aux effets de l'aide (Burnside et Dollar, 2000 ; Acemoglu et Weder, 2002 ; Acemoglu et *al.*, 2005) ; l'investissement (INV) ; le degré d'ouverture de l'économie (OUV) ; *un indicateur de stabilité macroéconomique* : le taux d'inflation (INFL) ; le déficit public en % du PIB (DEFP), et *un indicateur de capital humain* (KH) mesuré par le taux d'inscription à l'enseignement secondaire.

### 3.4. Les sources de données

Dans le cadre de cette étude qui couvre les pays de la zone UEMOA, les données annuelles sur toutes nos variables sont extraites de plusieurs bases : d'abord, la première est le *Système de notation des pays créanciers (SNPC) de l'OCDE.* De celle-ci est extraite l'aide publique au développement globale nette reçue par les pays en développement. Ensuite, la seconde est l'*International Country Risk Guide (ICRG)* détenue par le « *Political Risk Services* », où sont extraites les variables institutionnelles. Une autre base dont nous nous sommes servis est le *World Development Indicateur (WDI)*. Cette dernière contient des données en série normalement longue sur les autres variables du modèle. L'étude couvre la période 1980-2015[6].

### 3.5. Analyse descriptive des variables du modèle

L'analyse des données de notre base nous a amené à identifier les régimes des économies de l'Union par rapport au seuil d'aide publique au développement. En effet, il a été possible de décomposer les différentes combinaisons de ces régimes. On peut ainsi identifier deux (02) différents états à partir desquels les pays de l'Union pourraient se situer : une situation dans laquelle l'aide est inférieure à sa valeur seuil et celle dans laquelle on observe que l'aide est supérieure à sa valeur seuil. A la suite de cette identification, nous présentons alors la statistique descriptive des relations linéaire puis non linéaire l'aide et la croissance économique au sein de la Zone. Les résultats de ces analyses sont compilés dans le tableau 1, duquel plusieurs enseignements peuvent également être tirés.

D'abord, on constate que chaque régime contient au moins 42% du total des observations. Ainsi, l'inférence statistique et économétrique est applicable car chaque régime dispose assez de données pour obtention des tests concluants et des estimations assez cohérentes. Ensuite, sans aller dans les détails, la moyenne d'aide est 8.17% dans le premier régime conte 25.81% dans le second. Il y a bien une différence fondamentale entre les deux régimes. Enfin, le taux de croissance moyen n'est pas relativement stable dans les deux régimes. Il varie de 2.95% à 3.59%. Après l'analyse de ces résultats, nous procédons aux tests statistiques et économétriques sur nos données.

---

[5] - Le choix de ces variables spécifiques a été fait de façon à minimiser le risque de corrélation avec les variables du modèle. Ainsi, nous étudions la matrice des corrélations entre les variables du modèle afin d'éviter les biais statistiques et de réduire les risques de multi colinéarité.

[6] - Test de robustesse : pour respecter les règles qui conditionnent l'application des GMM, nous avons calculé les moyennes quinquennales sur nos observations. Cette méthode a l'avantage non seulement d'éliminer la tendance cyclique mais également de rendre le *T* relativement plus petit que le *N* (Giovanni, 2005 ; Ciocchini, 2006).

**Tableau 1** : Statistiques descriptives des linéaires et de seuil entre les variables

| Variables | Linéaire | Non linéaire | |
|---|---|---|---|
| | | ≤ 12.741 | > 12.741 |
| $\overline{TCPIB}$ | 4.1316 | 2.9536 | 3.5953 |
| $\sigma_{TCPIB}$ | 9.8584 | 4.4078 | 5.3992 |
| $TCPIB_{min}$ | -36.045 | -36.045 | -28.099 |
| $TCPIB_{max}$ | 20.286 | 14.576 | 20.286 |
| $\overline{APD}$ | 12.038 | 8.1734 | 25.814 |
| $\sigma_{APD}$ | 11.715 | 3.2236 | 4.0509 |
| $APD_{min}$ | 0.0742 | 0.0742 | 12.854 |
| $APD_{max}$ | 78.707 | 12.740 | 78.707 |
| $\overline{INST}$ | 2.5853 | 1.4881 | 2.9631 |
| $\sigma_{INST}$ | 1.4442 | 2.4192 | 2.9608 |
| $INST_{min}$ | 0.9421 | 0.9421 | 2.0922 |
| $INST_{max}$ | 3.5213 | 3.5201 | 3.5213 |
| $\overline{INV}$ | 19.463 | 19.338 | 17.276 |
| $\sigma_{INV}$ | 9.0727 | 7.0755 | 6.7031 |
| $INV_{min}$ | 2.7328 | 2.7328 | 4.5624 |
| $INV_{max}$ | 60.156 | 40.268 | 60.156 |
| $\overline{OUV}$ | 82.368 | 76.288 | 59.138 |
| $\sigma_{OUV}$ | 4.3183 | 6.4385 | 2.4515 |
| $OUV_{min}$ | 9.7830 | 25.916 | 9.7830 |
| $OUV_{max}$ | 171.71 | 171.71 | 138.27 |
| $\overline{INFL}$ | 5.7282 | 3.6108 | 11.224 |
| $\sigma_{INFL}$ | 11.640 | 4.4021 | 19.442 |
| $INFL_{min}$ | -17.641 | -17.641 | -7.7966 |
| $INFL_{max}$ | 80.788 | 22.913 | 80.788 |
| $\overline{DEFP}$ | -3.9295 | -3.6794 | -5.9901 |
| $\sigma_{DEFP}$ | 3.8375 | 3.5042 | 4.0039 |
| $DEFP_{min}$ | -14.647 | -14.647 | -3.7341 |
| $DEFP_{max}$ | 11.532 | -2.1643 | 11.532 |
| $\overline{KH}$ | 24.579 | 25.337 | 14.361 |
| $\sigma_{KH}$ | 15.378 | 12.872 | 9.6181 |
| $KH_{min}$ | 2.4843 | 2.4843 | 4.5137 |
| $KH_{max}$ | 72.313 | 55.911 | 72.313 |
| N | 288 | 165 | 123 |

**Source** : Auteur (2017), outputs du logiciel STATA

*Note: $\overline{X}$ représente les moyennes respectives correspondantes aux variables X. $X_{min}$ et $X_{max}$ indiquent les valeurs minimales et maximales. $\sigma_X$ est l'écart type et N le nombre d'observation.*

### 3.6. Résultats de l'inférence statistique et économétrique

Dans ce paragraphe, on présent les résultats des tests statistiques et économétriques que nous avons appliqué avant l'estimation des modèles à effet de seuil. Il s'agit notamment des tests de corrélation et de stationnarité d'une part et des tests de linéarité, de détermination du nombre de régimes et de localisation de l'intervalle de confiance sur le (s) seuil (s) respectif (s) déterminé (s) d'autre part. A cet effet, les résultats des tests de multicolinéarité et de stationnarité en données de panel (tests de Harris et Tzavalis, 1999 ; de Hadri, 2000 ; de Levin, Lin et Chu, 2002 ; et de Im, Pesaran et Shin, 2003) sont présentés ci-dessous. Pendant que les premiers révèlent l'existence d'une faible corrélation entre les variables du modèle, signifiant que l'inclusion de toutes ces variables dans un même modèle (équation) ne posera aucun problème de multicolinéarité ; les seconds ont révélé essentiellement que les variables : taux de croissance du PIB, l'aide publique au développement et l'investissement sont stationnaires avec l'introduction d'un trend dans le modèle. Toutefois, compte tenu de la puissance de test IPS (Im, Pesaran et Shin, 2003) sur les autres tests (Hurlin et Mignon, 2005), la conclusion retient globalement une absence de racines unitaires pour ces séries en panel.

**Tableau 2** : Matrice de corrélation des variables incluses dans le modèle

| Variables | TCPIB | APD | INST | INV | OUV | INFL | DEFP | KH |
|---|---|---|---|---|---|---|---|---|
| TCPIB | 1.000 | | | | | | | |
| APD | -0.026 | 1.000 | | | | | | |
| INST | 0.072 | -0.159 | 1.000 | | | | | |
| INV | 0.126 | 0.024 | 0.081 | 1.000 | | | | |
| OUV | 0.091 | 0.539 | -0.056 | -0.014 | 1.000 | | | |
| INFL | 0.038 | 0.491 | -0.104 | -0.499 | -0.178 | 1.000 | | |
| DEFP | 0.522 | 0.378 | -0.306 | 0.354 | 0.405 | 0.070 | 1.000 | |
| KH | 0.065 | 0.367 | 0.108 | 0.391 | 0.539 | 0.191 | 0.491 | 1.000 |

**Source :** Auteur (2017), outputs du logiciel STATA

A présent, il convient de présenter les résultats de l'inférence statistique du modèle dont les aspects théoriques sont clairement présentés plus haut.

A cet effet, dans cet article, nous utilisons l'algorithme de détermination de seuil endogène fourni par Hansen (1999) à l'exemple de Hurlin et villieu (2010), de Fiodendji et Evlo (2013) et de Fiodendji et *al.* (2014)[7]. Il s'agit d'une procédure de régression basée sur la technique des *moindres carrées séquentiels* sur toutes les valeurs seuils candidates jusqu'à ce que l'on obtienne la valeur seuil c'est à dire le seuil optimal correspond à la valeur de $\gamma$ qui minimise la somme des carrés des résidus. Après, le test de linéarité du modèle, de détermination du nombre de régimes et de l'estimation de l'intervalle de confiance du seuil optimal sont effectués sur la base des recommandations de Hansen relatives à l'utilisation du test de ratio de vraisemblance et de

---

[7] - Voir la programmation initiale de Hansen (1999) sur « //www.ssc.wisc.edu/~bhansen/ ». Elle a été reprise par Colletaz et Hurlin C. (2006, 2011), Fiodendji K.et Evlo K. (2013), et Fiodendji K., Kamgnia D. B. et Tanimoune, Ary N. (2014). La version modifiée du programme permet même de conditionner la détermination du seuil d'une variable par une autre variable d'intérêt préalablement définie.

la procédure de *bootstrap* (Hansen, 1996 ; 1999). Ainsi, la compilation minutieuse des résultats de ces tests se retrouve dans le tableau ci-dessous (tableau 4).

**Tableau 3** : Résultats de tests de racine unitaire sur données de panel

| Variables | TCPIB | APD | INST | INV | OUV | INFL | DEFP | KH |
|---|---|---|---|---|---|---|---|---|
| **Intercept** | | | | | | | | |
| Levin, Lin et Chu | -7.489[a] | -2.638[b] | -9.402[a] | -2.023[c] | -2.739[b] | -8.313[b] | -5.368[b] | -3.005[b] |
| | (0.002) | (0.026) | (0.001) | (0.084) | (0.019) | (0.043) | (0.029) | (0.042) |
| Harris et Tzavalis | -27.77[b] | -4.975[a] | -10.86[b] | -5.381[c] | -18.65[a] | -16.93[c] | -1.08 | -4.308[b] |
| | (0.020) | (0.000) | (0.037) | (0.084) | (0.001) | (0.076) | (0.219) | (0.039) |
| Im, Pesaran et Shin | -4.002[a] | -3.709[a] | -6.208[a] | -2.129[b] | -4.285[a] | -9.203[a] | -4.029[b] | -3.854[b] |
| | (0.001) | (0.001) | (0.000) | (0.035) | (0.001) | (0.000) | (0.029) | (0.026) |
| Hadri | 3.238[b] | 46.51[a] | 7.953 | 38.19[c] | 40.59[a] | 14.69 | -9.812[b] | 18.81[c] |
| | (0.031) | (0.000) | (0.412) | (0.904) | (0.000) | (0.193) | (0.035) | (0.087) |
| **Intercept + trend** | | | | | | | | |
| Levin, Lin et Chu | -6.471[a] | -2.549[a] | -9.864[a] | -9.332[c] | -4.185[a] | -8.343[a] | -6.009[a] | -7.401[a] |
| | (0.000) | (0.001) | (0.000) | (0.082) | (0.003) | (0.000) | (0.002) | (0.001) |
| Harris et Tzavalis | -15.36[a] | -6.164[a] | -9.247[a] | -4.236[c] | -9.524[a] | -16.42[b] | -11.49[b] | -13.08[a] |
| | (0.000) | (0.000) | (0.000) | (0.094) | (0.000) | (0.014) | (0.037) | (0.002) |
| Im, Pesaran et Shin | -4.078[a] | -6.392[a] | -7.613[a] | -5.552[b] | -6.311[a] | -10.26[a] | -9.137[a] | -6.068[a] |
| | (0.000) | (0.000) | (0.002) | (0.027) | (0.000) | (0.000) | (0.000) | (0.001) |
| Hadri | 15.83[a] | 20.28[a] | 13.05[a] | 24.06[c] | 15.38[a] | 14.235[b] | 21.04[a] | 18.24[a] |
| | (0.000) | (0.000) | (0.002) | (0.098) | (0.000) | (0.029) | (0.003) | (0.002) |

**Source** : Auteur (2017), outputs du logiciel STATA

<u>Notes</u> : *Les valeurs entre parenthèses représentent les p-values ; **a, b** et **c** représentent respectivement la significativité à **1%, 5%** et à **10%**. L'ordre maximal de retard est fixé à trois (03). Le choix de cet ordre de retard est fait la méthode AIC. La méthode d'estimation spectrale utilisée est celle de Barlett.*

L'analyse de ce tableau révèlent que lorsqu'on applique le test de Hansen (1999) avec 1000 réplications du *bootstrap* pour tester la non-linéarité, on trouve que les ratios de vraisemblance de *LR Hansen Test* et ses *p-values* conduisent au rejet de l'hypothèse nulle au seuil critique de 5 %. Ce résultat traduit le fait qu'il existe une relation non-linéaire entre l'aide et la croissance économique des pays de l'UEMOA. Un tel résultat implique que soit déterminé le nombre de régimes du processus. A cet effet, le même tableau montre que le test itératif de Fisher (*Fisher Test*) permet de conclure que l'hypothèse nulle est acceptée pour un seuil critique de 5 % pour la période et les sous-périodes considérées. En conséquence, ils existent deux régimes d'aide. Ce résultat traduit l'idée selon laquelle la non-linéarité de la relation aide et croissance dans la Zone l'UEMOA donne lieu à la détermination d'une seule valeur seuil.

Dans cette procédure, nous trouvons que la valeur seuil d'APD qui minimise la somme des carrés des résidus issus des estimations des moindres carrés séquentiels est de 12.74 % du PIB dans l'UEMOA. L'intervalle de confiance, calculé sur la base de la distribution simulée par méthodologie de Hansen (1999)[8] indique qu'à un risque de première espèce de 5%, cette valeur seuil d'APD se logerait entre 10.43 % et 15.81 %. Ces résultats sont assez riches et intéressants surtout lorsqu'on les compare à ceux des autres études. En conséquence, ils sont relativement plus proches de celles de Boone (1996), de Clemens et *al*., (2004) et d'Alia et Anago (2014) déjà évoqués dans la littérature. Par contre ils restent moins proche de celles qu'on l'on retrouve dans les travaux de Kalyvitis et *al.* (2012) et de Fiodendji et Evlo (2013).

---

[8] - Dans cette étude, nous avons utilisé GAUSS. C'est en effet est un logiciel qui permet d'écrire un langage de programmation pour les statistiques et l'économétrie. Il est développé et vendu par Aptech *Systems*, USA.

Deux arguments peuvent être avancés pour justifier notre résultat. Le premier est relatif au contexte des études et renvoie au fait que ces travaux ne sont pas réalisés sur la même période. Le deuxième a trait à la méthodologie de détermination des seuils endogènes. C'est l'argument selon lequel la valeur du seuil d'aide pourrait varier non seulement selon le nombre de réplications du *bootstrap* appliqué, mais aussi et surtout selon que les simulations soient conditionnées à d'autres variables ou non.

**Tableau 4** : Résultats de l'inférence sur le modèle à effet de seuil en données de panel

| Zone | Zone UEMOA |
|---|---|
| Période | 1980-2015 |
| Seuils ($\gamma$) | 12.741 |
| IC | [10.438 - 15.809] |
| LR- Hansen Test | 31.075 |
| VC  10% | 25.063 |
| VC   5% | 27.381 |
| VC   1% | 32.754 |
| Fisher Test (p-value) | 3.285 |
|  | (0.239) |
| Bootstrap replications | 1000 |
| Bootstrap  p -value | (0.000) |

**Source** : Auteur (2017), outputs du logiciel GAUSS

A partir du graphique (graph 3), nous analysons le positionnement des pays de l'Union par rapport au seuil déterminé. Au premier constat, la valeur moyenne de l'aide reçue par les pays de l'Union reste inférieure au seuil estimé. Ce qui signifie qu'en moyenne, les Etats bénéficient moins d'assistance qu'ils devaient en bénéficier pour soutenir leur processus croissance économique. Ensuite, on constate clairement que la Guinée Bissau, le Mali et Niger se situent au-dessus du seuil alors que les autres pays de l'Union sont au-dessous.

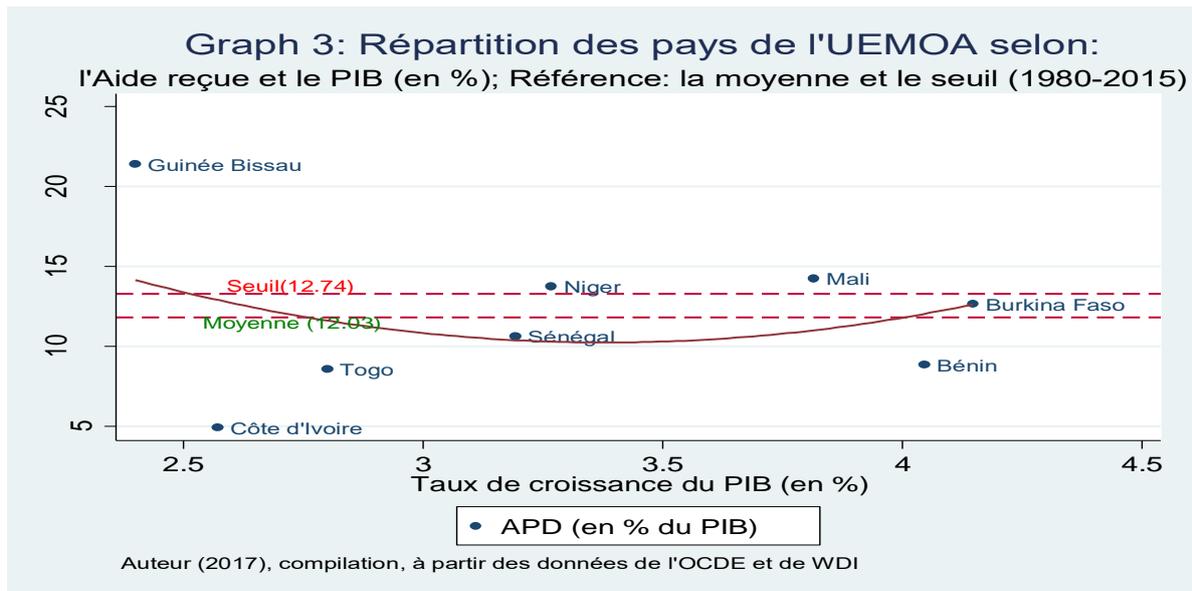

C'est un résultat qui est conforme aux faits stylés. Enfin, si l'on entre dans les détails, on remarque que la Guinée Bissau est le pays qui a le plus bénéficié de cette forme d'assistance alors la Côte d'Ivoire en est moins dépendante. Le Togo aussi en a moins bénéficié pour raison de son histoire relative à la rupture de la coopération pendant plus d'une décennie. Le Burkina Faso se situe entre la moyenne et le seuil alors le Sénégal et le Bénin se rapproche de la valeur moyenne.

La prochaine étape évalue l'effet de l'aide sur la croissance économique (équation de seuil) suivant la procédure sus décrite dans la méthodologie.

### 3.7. La prise compte de l'endogénéité de l'aide internationale

L'enjeu central des travaux de recherche sur le sujet de l'efficacité de l'aide concerne la prise en compte de la problématique de l'endogénéité de l'aide. En effet, il apparaît évident que l'aide ne peut pas être considérée comme exogène. Deux catégories d'instruments sont utilisées à cette fin. La première catégorie concerne les variables retardées de l'aide et de la variable dépendante (ici, le taux de croissance du PIB). Cependant, ceux-ci se sont révélés faibles du point de vue de leur efficacité dans la présente étude (Stock et Yogo, 2005). La seconde catégorie d'instruments a non seulement rapport avec la prise en compte de la qualité des institutions des pays bénéficiaires, mais également avec ceux proposés par Tavarez (2003) revisitée par Brun et *al.* (2008), Chauvet et *al.* (2008) et Ebeke et Drabo (2011) : les aides et les dons globaux pondérés respectivement par l'inverse de la distance entre le pays donateur et le pays receveur de l'aide. L'idée qui sous-tend cette procédure est que le niveau d'aide reçu par un pays donné de la part de l'un des principaux donateurs est fortement dépendant de la proximité géographique et culturelle, les alliances politiques et du commerce bilatéral entre le pays donateur et le pays bénéficiaire (Ebeke et Drabo, 2011, Mallaye et Yogo, 2013). Le choix de ces instruments s'est basé sur le test de Sargan (1958).

### 3.8. Résultats des effets seuil de l'aide sur la croissance économique

Dans ce paragraphe, nous analysons enfin les résultats des effets de seuil de l'aide sur la croissance à travers un modèle (équation 15) qui spécifie un vecteur de deux (02) coefficients $\beta_1$ et $\beta_2$ indiquant les effets de l'aide dans les régimes avant et après le seuil et dont les résultats sont reportés dans le tableau 5.

**Tableau 5** : Résultat de l'effet seuil direct de l'aide sur la croissance économique

| Zone | Zone UEMOA | |
| --- | --- | --- |
| Variables | 2SLS | SGMM |
| C | 18.09 | 15.42 |
| | (0.221) | (0.518) |
| TCPIB (-1) | - | 0.513*** |
| | - | (0.000) |
| APD ($\beta_1$) | 0.519 | 0.402 |
| | (0.182) | (0.599) |
| APD ($\beta_2$) | 0.693*** | 0.768*** |
| | (0.001) | (0.000) |
| INST | 0.0591 | 0.231* |
| | (0.394) | (0.087) |
| INV | 0.659*** | 0.638*** |
| | (0.000) | (0.000) |
| OUV | 0.073*** | 0.694* |
| | (0.001) | (0.071) |
| INFL | -0.018 | -0.076 |
| | (0.617) | (0.902) |
| DEFP | 0.068* | 0.089* |
| | (0.082) | (0.064) |
| KH | 0.804** | 0.493*** |
| | (0.025) | (0.002) |
| RMSE | 0.846 | - |
| Prob ($\chi^2$) | 0.000 | - |
| $R^2$ | 0.607 | - |
| Prob Wald test | - | 0.000 |
| Prob Sargan 2 | | 0.019 |
| Prob AR( 2) | - | 0.419 |
| Observations | 210 | 288 |
| Nbre de pays | 8 | 8 |

**Source :** Auteur (2017), à partir des estimations dans STATA

***Note*** *: Les valeurs entre parenthèses sont des valeurs de probabilité (p-value) ;*
*\*\*\*, \*\* et \* représentant respectivement la significativité à 1%, à 5%, et à 10%.*

Toutefois, avant d'interpréter ces résultats, nous prêtons d'abord une attention particulière à la qualité des estimations. En effet, dans le cas de la méthode des *2SLS*, nous avons observé la valeur de la racine carrée de l'erreur quadratique moyenne (RMSE). Cette statistique est faible (0.84). Ce qui traduit une meilleure qualité explicative du modèle. Pour le test de robustesse, l'estimateur *GMM* en système a été préféré par rapport à celui du *GMM* en différence première en raison du fait que ce premier a prouvé sa robustesse suite aux simulations de *Monte Carlo* effectuées par Blundel et Bond (1998). Dans chacun des deux (02) méthodes d'estimation, c'est le choix des instruments qui a déterminé la qualité des estimateurs. En effet, dans le cas présent, la probabilité associée au test de Sargan a validé le choix des instruments au seuil de 5% au regard des conditions sur les moments théoriques et empiriques. De plus, la probabilité adjointe au test de Wald qui y est reportée permet de confirmer globalement la qualité de l'estimation. Ainsi, il est alors loisible d'interpréter les résultats.

Ils révèlent que le paramètre d'ajustement associé au retard de la variable dépendante est positif et significatif au seuil de 1%. En effet, celui-ci dépend bien positivement de son niveau passé. Les coefficients d'aide ont des signes et des degrés de significativité différents à travers les deux régimes. Par exemple, lorsque d'aide est à un niveau supérieur au seuil, son effet sur la croissance des pays de la Zone est positif et significatif (0.69 au seuil de 1 %), peut-être à travers un *effet de spilovers*. Par contre, pour des valeurs d'aide inférieur au seuil, l'effet marginal de l'aide sur la croissance est certes positif, mais non significatif.

Par ailleurs, quand on analyse les variables de contrôle, on remarque qu'elles ont toutes un effet positif, exception faite à l'inflation. Ainsi, quel que soit le niveau d'aide allouée aux économies de l'UEMOA, l'ouverture commerciale (significatif), l'investissement (significatif), le capital humain (significatif), la qualité des institutions (non significatif) et le déficit budgétaire (non significatif) agissent positivement sur la croissance.

### 4. Conclusion et recommandation de politiques

Depuis plusieurs décennies, les Etats membres de l'UEMOA font partie des pays considérés comme les plus grands bénéficiaires de l'aide internationale. Cependant, les études empiriques relatifs à la relation aide-croissance aboutissent souvent à des résultats controversés. De plus, les débats sur la thématique de l'efficacité de l'aide ne tarissent pas. Cet article se positionne non seulement dans ce débat, mais également, fait un retour à cette relation aussi controversée entre l'aide internationale et la croissance. En effet, il a été d'aborder la question d'une façon différente : établir dans le cas des pays de l'UEMOA, un lien non linéaire entre les flux d'aides reçues, les dynamiques des facteurs structurels et les dynamiques de croissance économique de 1980 à 2015. Pour atteindre cet objectif, l'étude a appliqué un *Panel Threshold Regression* (PTR) à la Hansen (1999) et prend en compte la question de l'endogénéité de l'aide internationale par l'utilisation des *2SLS* et des *GMM*. En conséquence, plusieurs résultats forts se dégagent de cette étude :

- les ratios de vraisemblance de *LR Hansen Test* et ses *p-values* simulés par du *bootstrap* traduisent l'existence d'une relation non-linéaire entre l'aide et la croissance économique des pays de l'UEMOA ;
- l'aide publique au développement n'affecte positivement et significativement la croissance économique que lorsqu'elle atteint un seuil endogène de 12.74 % du PIB moyen des pays de l'Union ;
- dans le second régime, une augmentation de 1% de l'aide publique au développement se traduit par un accroissement de 0.61% de la croissance, *toutes choses égales par ailleurs*. Par contre, dans le premier régime, l'effet de l'aide est non significatif.

Au regard de nos résultats empiriques, il est recommandé d'accroitre le volume de l'aide en destination de ces pays qui aspirent atteindre les objectifs du développement durable (ODD). De plus, dans ces pays, la transparence dans la gestion de l'aide devra être améliorée.

Ces deux (02) premières recommandations interpellent à la fois les organismes ou pays donateurs et les pays bénéficiaire dans leurs Systèmes Nationaux de Gestion de l'Information sur l'Aide (SGIA) qui, devront instaurer une évaluation annuelle des performances des dispositifs d'aide et mobiliser surtout l'opinion publique sur le thème de l'efficacité de l'aide.

En définitive, nos pays doivent reconnaître que l'atteinte de leurs objectifs de croissance et développement socioéconomique devra être effectif plutôt par la mise en œuvre des politiques la mobilisation plus efficace des ressources domestiques qui peut être conditionnée par la mise place d'un système fiscal incitatif, que par la dépendance de l'aide étrangère quelle que soit sa nature, son contenu et sa destination. La Commission de l'UEMOA, pour sa part, devra désormais multiplier les plans d'urgence, les projets et programmes visant le développement socio-économique des pays de l'Union. Elle devra également procéder à un suivi rapproché dans l'application des textes communautaires au sein de ses Etats membres.

D'ailleurs, l'aide internationale est une ressource extérieure qui combine volatilité et incertitude. C'est pourquoi on ne cesse de se questionner sur son avenir. Ainsi, les études ultérieures gagneraient à analyser les impacts qu'aurait un choc (suppression) de l'aide internationale sur les économies bénéficiaires comme celles de l'UEMOA et les mécanismes d'ajustement qu'ils pourront formuler pour supporter un choc externe d'une telle nature.